\documentclass[twoside]{IEEEtran}

\usepackage{cite}
\usepackage[colorlinks=true, allcolors=blue]{hyperref}
\usepackage{booktabs}
\usepackage{graphicx}
\usepackage{stfloats}
\usepackage{amsmath}
\usepackage{amssymb}
\usepackage{soul}
\usepackage{xurl}
\usepackage{balance}
\usepackage{tabularx}
\usepackage{multirow}
\usepackage{threeparttable}
\usepackage{makecell}
\usepackage{subcaption}
\usepackage{tikz}
\usepackage{placeins}
\usetikzlibrary{arrows.meta, positioning, shapes.geometric, fit}

\DeclareRobustCommand*{\IEEEauthorrefmark}[1]{%
  \raisebox{0pt}[0pt][0pt]{\textsuperscript{\footnotesize #1}}%
}
\begin{document}

\title{{Design and Quantitative Evaluation of an Embedded EEG Instrumentation Platform for Real-Time SSVEP Decoding}}

\author{Manh-Dat~Nguyen\IEEEauthorrefmark{1},~\IEEEmembership{Student Member,~IEEE,}
Thomas~Do\IEEEauthorrefmark{1,}\IEEEauthorrefmark{*}, 
Nguyen Thanh Trung~Le\IEEEauthorrefmark{1},
Xuan-The~Tran\IEEEauthorrefmark{1},
Fred Chang\IEEEauthorrefmark{1},
and ~Chin-Teng~Lin\IEEEauthorrefmark{1}
\thanks{This work was supported by Computational Intelligence and Brain-Computer Interfaces lab, University of Technology Sydney, Australia. \textit{(Corresponding author: Thomas Do.)}}
\thanks{\IEEEauthorrefmark{1}School of Computer Science, Faculty of Engineering and Information Technology, University of Technology Sydney, Sydney, New South Wales, 2007, Australia (e-mail: 
 thomas.do@uts.edu.au).}
}

\markboth{TBA }{Manh Dat \MakeLowercase{\textit{et al.}}: Design and Quantitative Evaluation of an Embedded EEG Instrumentation Platform for Real-Time SSVEP Decoding}
\maketitle

\begin{abstract}
    This paper presents an embedded EEG instrumentation platform for real-time steady-state visually evoked potential (SSVEP) decoding based on an ESP32-S3 microcontroller and an ADS1299 analog front end. The system performs $8$-channel EEG acquisition, zero-phase bandpass filtering, and canonical correlation analysis entirely on-device, while supporting wireless communication and closed-loop operation without external computation. A central contribution is the quantitative characterization of the platform's measurement integrity. Reported results demonstrate a stable shorted-input noise floor ($\approx 0.08~\mu\text{V}_{\text{RMS}}$), tightly bounded sampling jitter ($0.56~\mu\text{s}$ standard deviation), and negligible long-term drift ($< 1~\text{ppm}$). Numerical fidelity analysis shows $100\%$ decision agreement between the mixed-precision embedded pipeline and a $64$-bit double-precision reference. Effective common-mode attenuation exceeded $112~\text{dB}$ under balanced conditions, with a localized $26.9~\text{dB}$ degradation observed under source-impedance mismatch. Closed-loop validation achieved $99.17\%$ online accuracy and an information transfer rate of $27.66~\text{bits/min}$. These results position the proposed system as a quantitatively characterized embedded EEG measurement and processing platform for real-time SSVEP decoding.
\end{abstract}

\section{Introduction}
\label{sec:introduction}

\IEEEPARstart{B}{rain}--computer interfaces (BCIs) enable direct communication between neural activity and external devices, with important applications in assistive technology, neurorehabilitation, and human--machine interaction~\cite{patrick2025state,willsey2025high,zhuang2025aegis,lin2020direct,do2020increase}. Among non-invasive BCI paradigms, steady-state visually evoked potentials (SSVEPs) are particularly attractive because of their relatively high signal-to-noise ratio, frequency-specific structure, and suitability for lightweight real-time decoding~\cite{cao2025novel,nakanishi-2017,angrisani-2023}. These properties make SSVEPs a strong candidate for portable and embedded BCI systems, especially in the broader context of on-device and Edge-AI-enabled neurotechnology~\cite{manhdat-2025}.

Despite this suitability, many practical SSVEP systems still rely on laboratory-grade acquisition hardware and host-side computation. Prior work has demonstrated FPGA-assisted systems, ultra-low-power wearable processors, and microcontroller-based implementations~\cite{lin-2020,kartsch-2019,teversham-2022}. However, most such studies primarily emphasize classification accuracy, information transfer rate, portability, or energy efficiency, rather than systematic instrumentation-level characterization of the embedded sensing and processing chain~\cite{arpaia2020wearable,arpaia-2020-ar-bci,xu2026wearable}. Comparatively fewer works evaluate the embedded system as a measurement and processing instrument through quantitative characterization of front-end noise, sampling-timing integrity, arithmetic fidelity, processing latency, or practical common-mode interference rejection.

This gap is important because the performance of portable EEG/SSVEP devices depends not only on the decoding algorithm, but also on the reliability of the full embedded signal chain~\cite{sun2025signal,liang2023development}. In a microcontroller-class implementation, design choices in sampling architecture, arithmetic precision, memory placement, communication load, and analog front-end configuration can directly affect timing determinism, numerical stability, interference sensitivity, and overall system robustness. For this reason, embedded SSVEP platforms should be evaluated not only by task-level decoding performance, but also by quantitative characterization of the measurement and processing chain.

This paper presents an embedded EEG instrumentation platform for real-time SSVEP decoding based on an ESP32-S3 microcontroller and an ADS1299 analog front end. The system acquires 8-channel EEG, performs zero-phase bandpass filtering and canonical correlation analysis (CCA) entirely on-device, and supports wireless communication and closed-loop operation without external computation. Beyond demonstrating embedded real-time decoding, the paper emphasizes quantitative characterization of the platform as an integrated measurement and processing system under both baseline and streaming-enabled operating conditions.

The main contributions of this work are as follows:
\begin{enumerate}
    \item \textbf{Embedded multichannel SSVEP platform}: Development of a fully embedded 8-channel EEG/SSVEP system that performs acquisition, zero-phase filtering, and CCA-based classification entirely on a microcontroller-class platform while supporting real-time wireless communication and closed-loop operation.
    
    \item \textbf{Measurement-oriented characterization}: Quantitative characterization of the embedded platform in terms of shorted-input noise repeatability, sampling jitter and drift, numerical fidelity relative to a double-precision reference, processing-latency distribution, effective common-mode attenuation, and power and memory footprint.
    
    \item \textbf{Closed-loop validation}: Validation in a six-target closed-loop SSVEP experiment, demonstrating 99.17\% online accuracy and practical real-time operation using the same deployed embedded pipeline characterized in the measurement results.
\end{enumerate}

The remainder of this paper is organized as follows. Section~\ref{sec:related_work} reviews related work on embedded and portable EEG/SSVEP systems. Section~\ref{sec:platform} describes the hardware architecture, firmware organization, embedded signal-processing pipeline, and communication interface. Section~\ref{sec:methodology} presents the experimental methodology and characterization procedures. Section~\ref{sec:characterization_validation_results} reports the quantitative characterization and closed-loop validation results. Section~\ref{sec:discussion} discusses the system-level implications, comparison with prior work, and remaining limitations. Finally, Section~\ref{sec:conclusion} concludes the paper.

\section{Related Work}
\label{sec:related_work}

This section reviews prior work on embedded SSVEP/EEG systems and highlights the main instrumentation-oriented gap addressed in this study.

\subsection{Embedded EEG and SSVEP Decoding}
\label{subsec:embedded_eeg_ssvep_systems}

SSVEP-based BCIs exploit frequency-specific neural responses to visual stimulation, typically strongest over occipital regions \cite{vialatte-2009}. Although methods such as Filter Bank CCA (FBCCA) \cite{chen-2015}, Task-Related Component Analysis (TRCA) \cite{nakanishi-2017}, and deep learning models such as EEGNet \cite{lawhern-2018} can improve decoding performance, their training requirements and computational cost make them less suitable for microcontroller-class deployment under tight memory and power constraints. For this reason, standard Canonical Correlation Analysis (CCA) \cite{lin-2006} remains a practical training-free approach for embedded SSVEP decoding. Commercial portable platforms such as Mentalab Explore, OpenBCI Cyton, and Muse S \cite{mentalab_explore, openbci_cyton, muse_s} mainly function as acquisition devices and typically rely on external hosts for downstream decoding. Fully embedded systems that combine multichannel EEG acquisition with real-time on-device SSVEP processing remain comparatively limited \cite{wang-2020}.

\subsection{Instrumentation Gaps in Embedded BCI Studies}
\label{subsec:measurement_gaps_prior_embedded_bci}

A central challenge in embedded BCI design is maintaining deterministic real-time operation under limited compute, memory, and communication resources \cite{wang-2020}. In addition, interoperability is often constrained because standard frameworks such as Lab Streaming Layer (LSL) \cite{kothe-lsl} are rarely implemented natively on microcontrollers due to resource overhead, leading many platforms to adopt simplified communication schemes or host-assisted workflows \cite{lin-2020, acampora-2020}.

Prior studies have primarily emphasized task-level outcomes such as classification accuracy, information transfer rate, portability, low power, or cost. By contrast, systematic instrumentation-oriented characterization remains limited, particularly for shorted-input noise repeatability, sampling jitter and drift, numerical fidelity of reduced-precision pipelines, processing-latency determinism, and practical common-mode rejection robustness. As a result, many embedded BCI systems are evaluated mainly as decoders rather than as quantitatively characterized EEG measurement and processing instruments.

This paper addresses that gap by combining real-time on-device SSVEP decoding with measurement-oriented characterization of signal quality, timing integrity, numerical fidelity, execution latency, and system-level common-mode attenuation.

\section{Embedded EEG Instrumentation Platform}
\label{sec:platform}

This section describes the proposed embedded EEG instrumentation platform, including the hardware, firmware, signal-processing pipeline, and communication interface used for real-time on-device SSVEP decoding.

\subsection{System Overview}
\label{subsec:system_overview}

\begin{figure*}[ht]
\centering
\includegraphics[width=\linewidth]{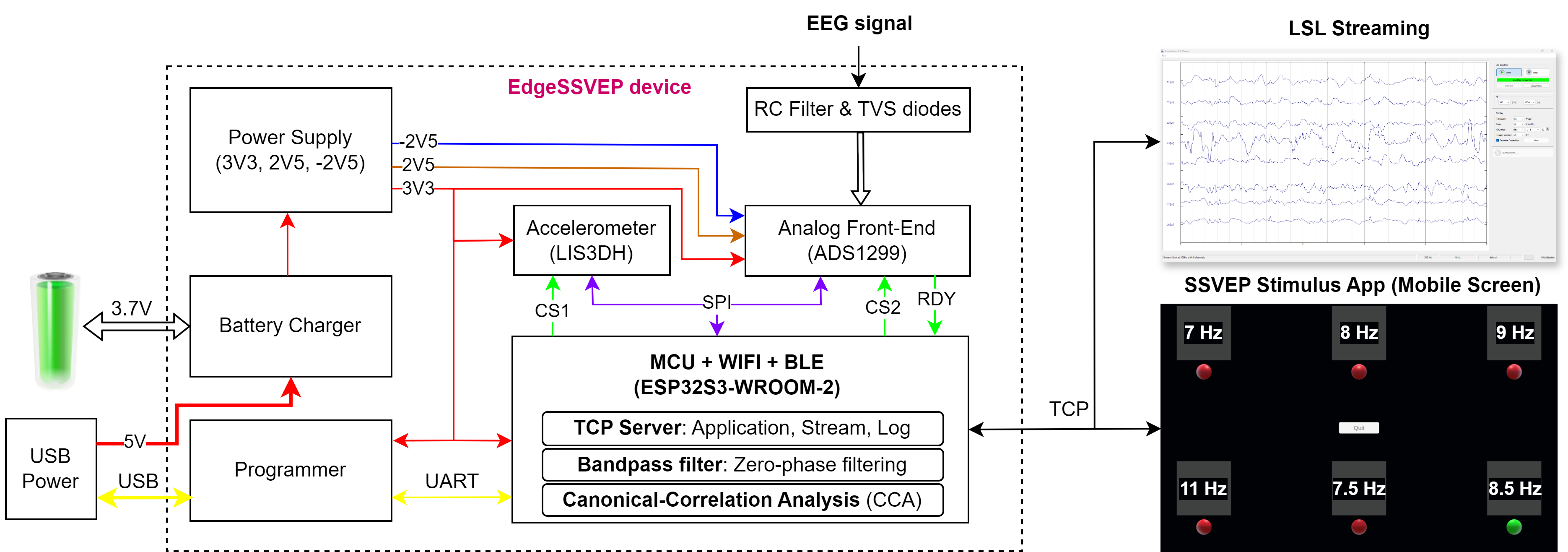}
\caption{Overall architecture of the embedded EEG instrumentation platform. The system integrates battery-powered operation, an ADS1299-based analog front end, an ESP32-S3 microcontroller for on-device zero-phase filtering and CCA-based SSVEP decoding, and a lightweight TCP interface for external communication.}
\label{fig:system_overview}
\end{figure*}

The proposed platform integrates EEG acquisition, on-device preprocessing, SSVEP decoding, and external communication in a battery-powered microcontroller-based device, as illustrated in Figure~\ref{fig:system_overview}. An ADS1299 acquires 8-channel EEG, while an ESP32-S3 performs embedded signal conditioning and CCA-based classification and exports data through a lightweight TCP interface for visualization, logging, or host-side integration. By separating timing-critical acquisition and processing from communication tasks, the platform supports continuous real-time operation in a compact form factor.

\subsection{Hardware Architecture}
\label{subsec:hardware_architecture}

The platform is built around an ESP32-S3 microcontroller and an ADS1299 analog front end. The ESP32-S3 provides dual-core embedded processing for acquisition control, filtering, classification, and telemetry, while the ADS1299 supports synchronized 8-channel biopotential acquisition with 24-bit resolution and programmable gain. The device is operated at 500~SPS, which is sufficient for the targeted SSVEP range while supporting stable zero-phase filtering and CCA-based recognition.

Table~\ref{tab:hardware-specs} summarizes the main hardware features. The platform integrates the EEG front end, microcontroller, external memory, wireless communication, accelerometer, and battery-powered operation on a single PCB, enabling self-contained EEG acquisition and on-device decoding without external computation.

\begin{table}[t]
\centering
\caption{Summary of key hardware specifications.}
\label{tab:hardware-specs}
\begin{tabular}{p{3.1cm}p{4.9cm}}
\toprule
\textbf{Component} & \textbf{Specification} \\
\midrule
Microcontroller & ESP32-S3-WROOM-2, dual-core Xtensa LX7, up to 240\,MHz \\
External memory & 32\,MB Flash, 16\,MB PSRAM \\
Analog front end & ADS1299, 8-channel, 24-bit resolution \\
Sampling configuration & 8 channels at 500\,SPS, PGA = 12 \\
Wireless interface & Wi-Fi and BLE \\
Motion sensor & LIS3DH, 3-axis digital accelerometer \\
Power supply & 3.7\,V battery or USB \\
Protection network & RC input network and TVS diodes \\
PCB & 4-layer stack-up, 65\,mm $\times$ 55\,mm \\
\bottomrule
\end{tabular}
\end{table}

\subsection{Analog Front-End and Input Protection Network}
\label{subsec:analog_frontend_input_protection}

The analog front end is designed to preserve measurement fidelity during wearable operation while protecting the acquisition circuitry from handling disturbances. Each input uses a 4.99~k$\Omega$ series resistor and a 4.7~nF capacitor for high-frequency interference attenuation and input protection. TVS diodes are placed at the connector interface to reduce susceptibility to electrostatic discharge during electrode attachment and removal. The analog and digital rails are derived from low-noise battery-powered supplies, preserving the same front-end signal path used in the subsequent characterization measurements.

\subsection{Real-Time Firmware Architecture}
\label{subsec:firmware_architecture}

The firmware follows a modular FreeRTOS architecture that separates timing-critical acquisition and processing from communication and logging. A hardware-driven interrupt service routine (ISR) monitors the ADS1299 data-ready (DRDY) signal and immediately notifies the acquisition task, minimizing scheduling latency and supporting deterministic sampling.

To support later timing characterization, the firmware records microsecond-resolution device timestamps at key pipeline boundaries, including DRDY events and processing-stage start and end points. To avoid contaminating the measured compute interval with communication overhead, timing records are enqueued after the timed region and transmitted asynchronously by a dedicated logging task.

\begin{figure}[!t]
\centering
\includegraphics[width=0.9\linewidth]{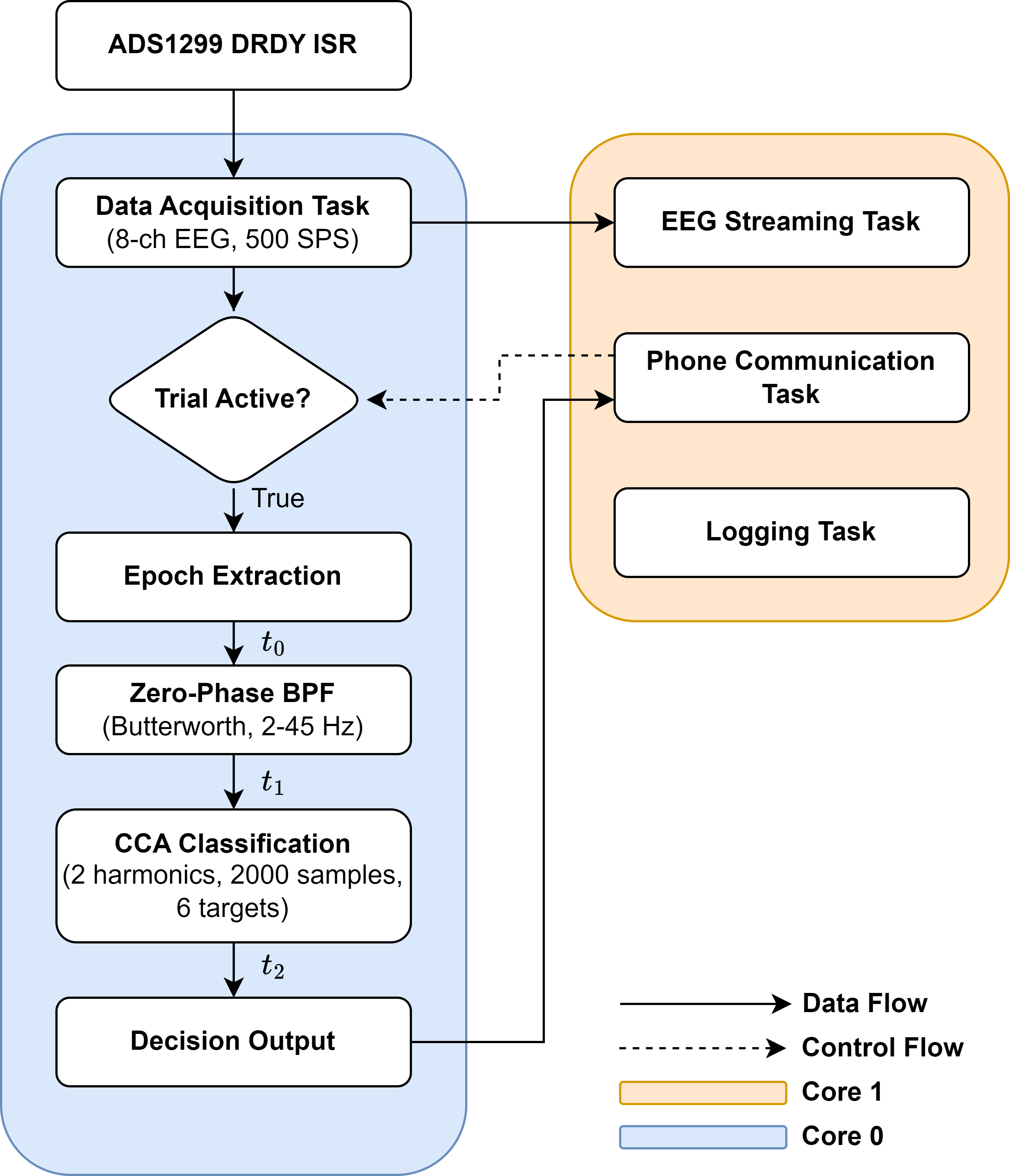}

\caption{Embedded firmware architecture. A hardware-driven ADS1299 DRDY interrupt service routine triggers the Core~0 data acquisition task. Core~0 performs epoch extraction, zero-phase bandpass filtering, and CCA-based classification only when the trial state is enabled through the phone communication path on Core~1. Core~1 also manages EEG streaming and optional logging. The decision output is returned through the communication task for real-time feedback display on the phone. The timestamp points \(t_0\), \(t_1\), and \(t_2\) denote the onset of filtering, the start of CCA classification, and the completion of the decision output, respectively.}
\label{fig:firmware-arch}
\end{figure}

As illustrated in Figure~\ref{fig:firmware-arch}, real-time operations execute on Core~0, including data acquisition, buffering, zero-phase bandpass filtering, epoch extraction, and CCA classification. Core~1 manages EEG streaming, phone communication, and logging. This separation prevents network delays or client-side latency from interfering with the sampling loop. Inter-task coordination relies on lightweight FreeRTOS direct task notifications and shared buffers, providing a low-latency path between acquisition, filtering, and classification while allowing continuous transmission on the communication core.

\subsection{Embedded Signal-Processing Pipeline}
\label{subsec:signal_processing_pipeline}

The embedded pipeline applies zero-phase bandpass filtering followed by CCA-based frequency recognition. The bandpass stage is implemented as a forward--backward IIR filter using a 3rd-order Butterworth design with a 2--45\,Hz passband. This suppresses slow drifts and high-frequency noise while preserving the oscillatory components and harmonics relevant to SSVEP decoding.

The filtering implementation is optimized for embedded deployment on the ESP32-S3. A Direct Form~II transposed structure is used to reduce memory overhead, and signal padding is applied to suppress boundary transients. In the deployed configuration, filtering is applied to the final 4-second segment of each 5-second trial recorded at 500~Hz, and the same filtered segment is then passed to the classifier. This matches the operating configuration used in closed-loop validation and is supported by the temporal-window comparison in Table~\ref{tab:window_comparison}.

CCA is used as a lightweight training-free method for real-time SSVEP recognition. In this implementation, six stimulation frequencies are considered, and each reference matrix includes two harmonics, yielding four basis functions per target frequency. For each candidate frequency, covariance and cross-covariance matrices are computed from the EEG segment and the corresponding reference signals. The resulting CCA problem is reformulated from a generalized eigenvalue problem into a standard eigenvalue problem by inverting the EEG covariance matrix \cite{hardoon-2004}. The dominant eigenvalue is then used to derive the canonical correlation coefficient, and the frequency associated with the maximum coefficient is selected as the classifier output.

To support deterministic embedded execution, the CCA module uses fixed memory allocation, single-precision arithmetic, and custom linear-algebra routines implemented in C. The processing chain includes zero-meaning of the EEG segment, covariance and cross-covariance computation, matrix inversion using Gauss--Jordan elimination, and dominant-eigenvalue extraction via power iteration \cite{allen2017doubly}. This allows real-time classification of 8-channel EEG data using only the onboard resources of the ESP32-S3.

\subsection{Communication Interface and Implementation Footprint}
\label{subsec:communication_interface_implementation_footprint}

A lightweight TCP communication layer supports continuous real-time streaming of EEG and auxiliary data. Each acquired sample is packed into a compact frame containing EEG channels, battery status, an event marker, and, when enabled, three-axis accelerometer readings, keeping all streamed data within the same device clock domain. Frames are buffered and transmitted asynchronously by a dedicated TCP task, decoupling network activity from the acquisition loop.

When compatibility with standard neuroscience toolchains is required, a host computer can receive the TCP stream and republish it through LSL. Interoperability with LSL-based workflows is therefore provided through host-side bridging rather than native on-device LSL support.

The signal-processing pipeline also uses explicit memory placement to balance runtime efficiency and embedded resource constraints. The zero-phase filter uses statically allocated work buffers, the CCA module stores large sample and transpose buffers in PSRAM, smaller covariance and inverse matrices remain in internal SRAM for faster access, and the precomputed sinusoidal reference bank is stored in read-only memory. These choices make the embedded filtering and classification pipeline practical on a microcontroller-class platform without external computation.

\section{Measurement Methodology}
\label{sec:methodology}

This section describes the experimental procedures used to characterize the embedded EEG platform from both instrumentation and application perspectives.

\subsection{Operating Modes and Test Conditions}
\label{subsec:operating_modes_test_conditions}

Unless otherwise stated, all characterization experiments were performed with the embedded platform configured for 8-channel single-ended EEG acquisition at 500~SPS, a programmable gain of 12, and an ESP32-S3 CPU frequency of 240~MHz. In this configuration, all EEG channels were acquired with respect to the common SRB1 reference, and the BIAS node was enabled as in normal system operation. These settings match the deployed real-time SSVEP operating configuration used throughout the embedded pipeline and the closed-loop validation experiments. The device was typically powered from its internal battery so that the reported measurements reflected normal standalone operation.

\begin{figure*}[ht]
\centering
\begin{subfigure}[t]{0.32\textwidth}
    \centering
    \includegraphics[width=\linewidth]{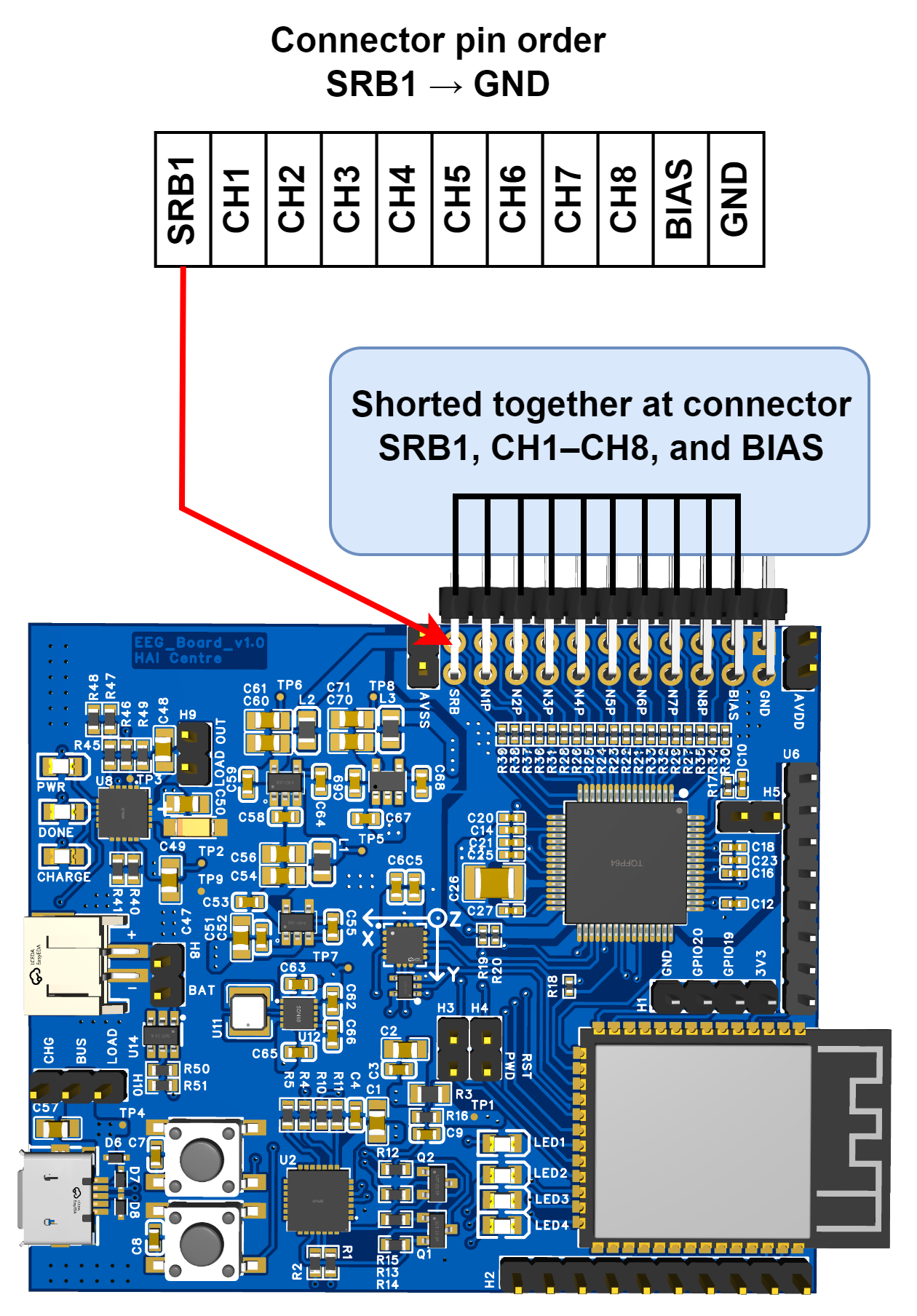}
    \caption{Shorted-input noise configuration.}
    \label{fig:method_noise_setup}
\end{subfigure}
\hfill
\begin{subfigure}[t]{0.32\textwidth}
    \centering
    \includegraphics[width=\linewidth]{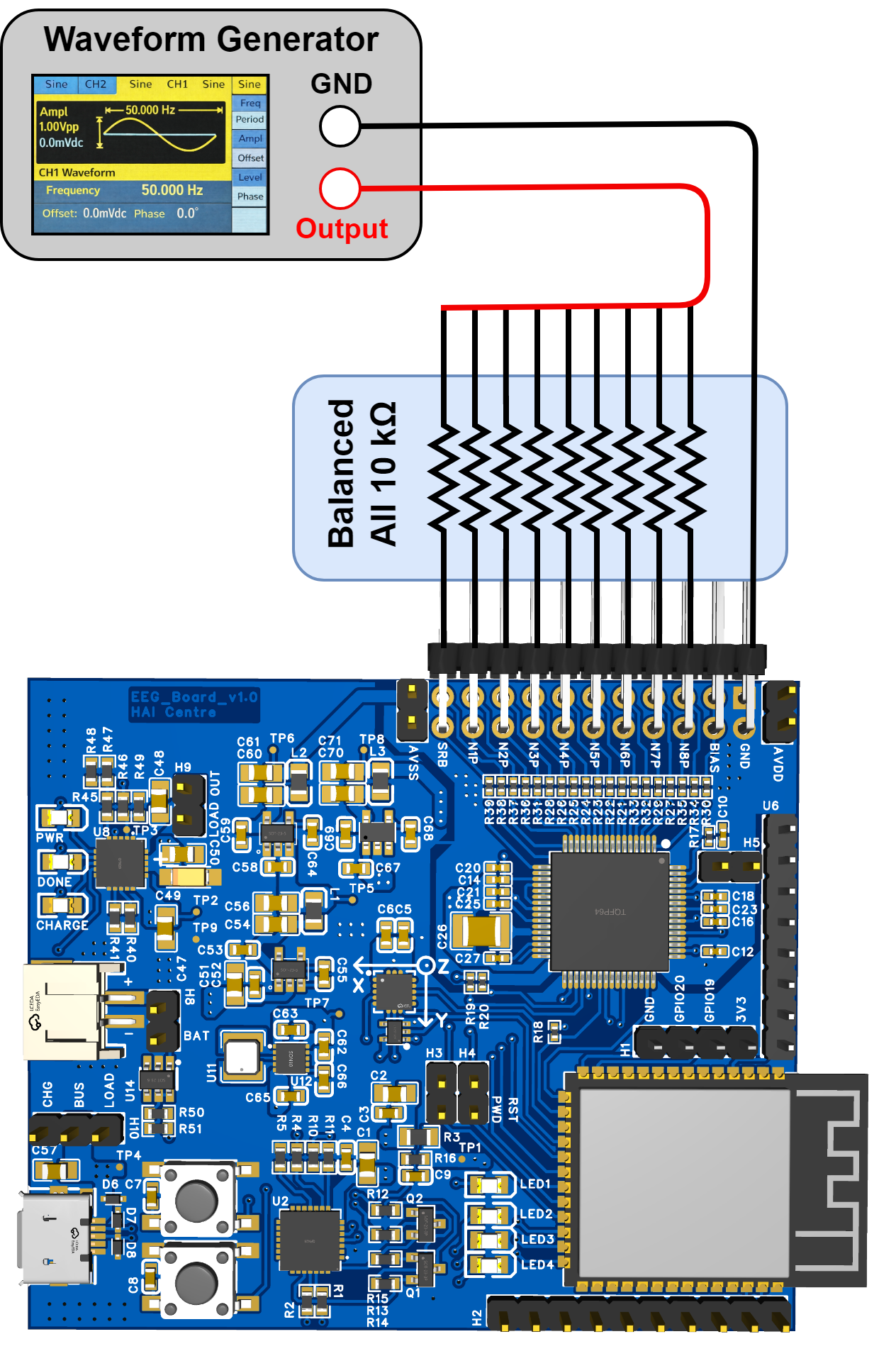}
    \caption{Effective common-mode rejection setup under the balanced condition.}
    \label{fig:method_cmrr_balanced}
\end{subfigure}
\hfill
\begin{subfigure}[t]{0.32\textwidth}
    \centering
    \includegraphics[width=\linewidth]{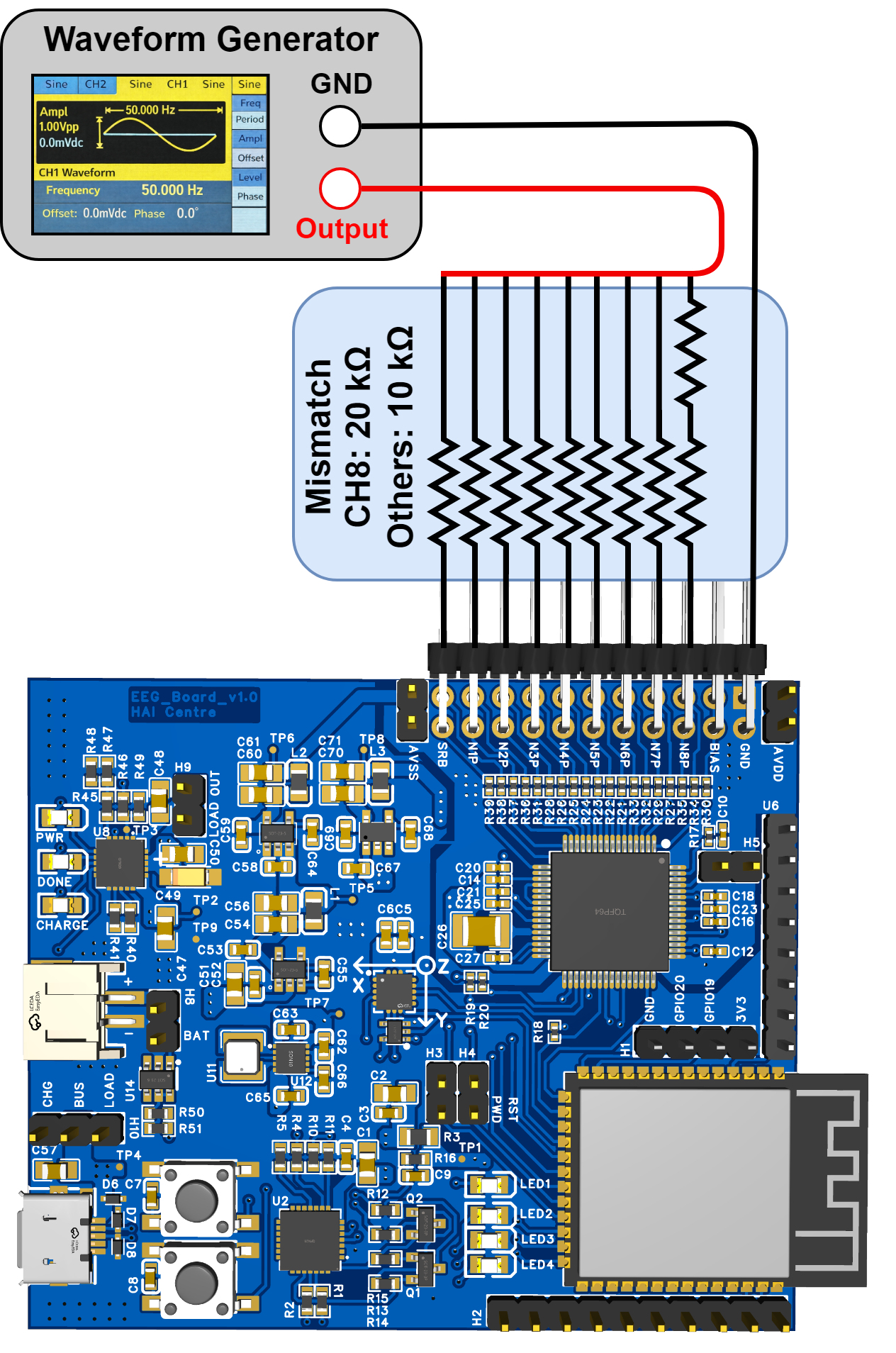}
    \caption{Effective common-mode rejection setup under the mismatch condition.}
    \label{fig:method_cmrr_mismatch}
\end{subfigure}
\caption{Measurement setups used for noise characterization and effective common-mode rejection evaluation.}
\label{fig:measurement_setups}
\end{figure*}

Two operating modes were considered throughout the characterization. In the \textit{OFF} mode, high-rate external EEG streaming was disabled and the embedded system operated in its baseline configuration, while maintaining the normal control path required for interaction with the external client or mobile application. In the \textit{ON} mode, continuous TCP-based raw-data streaming was enabled in addition to the normal embedded acquisition and processing pipeline, thereby representing a stress condition associated with sustained wireless telemetry during real-time operation. Thus, the OFF/ON distinction reflects the presence or absence of continuous high-rate streaming rather than a change in the core signal-processing algorithm. The measurement setups used for shorted-input noise evaluation and effective common-mode rejection testing are illustrated in Figure~\ref{fig:measurement_setups}.

\subsection{Noise Characterization}
\label{subsec:noise_characterization}

The shorted-input noise floor was evaluated to assess the system-level front-end noise under the same wiring topology used during operation, as illustrated in Figure~\ref{fig:method_noise_setup}. For this test, all eight channel inputs, SRB1, and the BIAS node were externally shorted together at the electrode connector. The short was applied before the RC input protection network so that the complete analog input path remained identical to normal operation while excluding electrode-interface effects.

Noise floor repeatability under communication load was assessed in both OFF and ON operating modes. For each mode, six runs were recorded, each consisting of a 30~second acquisition using the shorted-input configuration. Each run was segmented into six non-overlapping 5~second windows.

Two noise metrics were evaluated. The primary metric was EEG-band RMS noise, obtained after zero-phase Butterworth bandpass filtering from 2 to 45~Hz to match the passband of the embedded signal-processing pipeline used for real-time SSVEP decoding. A secondary metric was wideband RMS noise after DC removal only, without bandpass filtering. For each window, RMS noise was computed separately for each channel and then summarized as the mean across the eight EEG channels. The reported summary statistics were the mean and standard deviation across runs for each operating mode.

\subsection{Sampling Timing Characterization}
\label{subsec:sampling_timing_characterization}

Sampling timing integrity was evaluated by logging a microsecond-resolution device timestamp at every ADS1299 data-ready (DRDY) event. Timing records were generated on-device and transmitted asynchronously by a dedicated logging task so that the logging path did not perturb the acquisition timing under test. The recorded timing logs contained four fields: a sequential sample index, an absolute cumulative timestamp in microseconds, the corresponding time in seconds, and the operating mode label.

For each operating mode, a single continuous recording of 10~minutes was acquired at a nominal sampling rate of 500~Hz, corresponding to 300{,}000 events and an expected inter-sample period of 2000~\(\mu\)s. Inter-sample intervals were computed from the absolute timestamp sequence as
\[
\Delta t_i = t_i - t_{i-1}.
\]

Sampling jitter was characterized using the mean, standard deviation, median (P50), upper-tail percentiles (P95 and P99), and worst-case minimum and maximum values of \(\Delta t\). Long-term drift was estimated by partitioning the interval sequence into non-overlapping 10-second windows, computing the mean interval within each window, fitting a first-order linear model to the windowed means over elapsed time, and converting the fitted slope to parts per million relative to the expected period. This procedure was applied to both OFF and ON operating modes to assess the effect of sustained streaming activity on sampling timing integrity.

\subsection{Numerical Fidelity Evaluation}
\label{subsec:numerical_fidelity_evaluation}

Numerical fidelity was evaluated to quantify how closely the embedded mixed-precision processing pipeline reproduced a higher-precision offline reference. The same raw EEG dataset was replayed offline through four numerical configurations: double-precision filtering with double-precision CCA (DD), double-precision filtering with single-precision CCA (DF), single-precision filtering with single-precision CCA (FF), and single-precision filtering with double-precision CCA (FD). The DF configuration corresponds to the arithmetic used in the embedded implementation and was therefore treated as the software-equivalent representation of the deployed pipeline.

For each trial, the predicted stimulation class was recorded together with correlation-based confidence measures derived from the CCA output. In particular, the peak canonical correlation coefficient, denoted by \(\rho_{\mathrm{peak}}\), and the decision margin, defined as the difference between the largest and second-largest correlation coefficients, were extracted for each configuration. The DD configuration was used as the numerical reference for comparison.

Two complementary fidelity metrics were considered. The first metric was decision agreement, defined as the proportion of trials for which a given configuration produced the same predicted class as the DD reference. The second metric was the maximum absolute deviation in decision margin relative to DD, which was used to assess the sensitivity of confidence-related quantities to reduced numerical precision. This evaluation was intended to distinguish the effects of reduced precision in the filtering stage from those arising in the CCA stage.

\subsection{Processing Latency Characterization}
\label{subsec:processing_latency_characterization}

Processing latency determinism was characterized by instrumenting the on-device signal-processing pipeline with microsecond-resolution timestamps captured at three stage boundaries: the onset of zero-phase bandpass filtering (\(t_0\)), the end of filtering and start of CCA classification (\(t_1\)), and the completion of the CCA decision (\(t_2\)). Timing records were generated on-device and transmitted asynchronously by a dedicated logging task so that communication overhead was excluded from the timed region.

Three latency components were evaluated. The bandpass-filtering latency was defined as \(t_1 - t_0\), the CCA latency as \(t_2 - t_1\), and the total processing latency as
\[
t_{\mathrm{proc}} = t_2 - t_0.
\]

The evaluation was conducted under both OFF and ON operating modes. For each mode, 170 consecutive processing cycles were recorded over approximately 33 minutes. Each cycle corresponded to one trial-level decision, in which the final 4-second EEG segment from the 5-second trial was processed by the zero-phase bandpass filter and the CCA classifier. For each latency component, the reported statistics included the mean, sample standard deviation, median, 95th percentile, 99th percentile, and maximum value. This procedure was used to assess the effect of sustained streaming activity on the timing determinism of the embedded processing pipeline.

\subsection{Common-Mode Rejection Evaluation}
\label{subsec:common_mode_rejection_evaluation}

Effective common-mode rejection was evaluated at the system level using the balanced and mismatch common-mode injection setups shown in Figure~\ref{fig:method_cmrr_balanced} and Figure~\ref{fig:method_cmrr_mismatch}, respectively. A 50~Hz sinusoidal signal with a nominal amplitude of 1~V\(_{\mathrm{pp}}\), as set on a BK Precision 4052 function generator, was applied to a common injection node connected to Channels~1--8 and SRB1 through external resistors. The BIAS electrode was left floating and was not included in the injection path. To preserve the injected component for analysis, the recorded data were acquired without additional digital notch or bandpass filtering during capture.

Two source-impedance conditions were tested. In the balanced condition, all eight channels and SRB1 were connected to the common injection node through identical 10~k\(\Omega\) resistors. In the mismatch condition, Channels~1--7 and SRB1 remained connected through 10~k\(\Omega\) resistors, whereas Channel~8 was intentionally perturbed by increasing its series resistor to 20~k\(\Omega\). This condition was introduced to emulate source-impedance asymmetry and to examine its effect on channel-wise common-mode attenuation.

For each source-impedance condition, two recordings were acquired: one with the waveform generator disabled and one with the 50~Hz common-mode injection enabled. Thus, four recordings were obtained in total: balanced/OFF, balanced/ON, mismatch/OFF, and mismatch/ON. Each recording was approximately 30~seconds in duration. The generator-OFF recording for each source-impedance condition was used as its corresponding baseline reference, so that baseline correction was performed separately for the balanced and mismatch setups.

The 50~Hz component in each channel was extracted using a single-frequency discrete Fourier transform evaluated at exactly 50~Hz over the full recording length, yielding the root-mean-square (RMS) amplitude of the 50~Hz component. For each source-impedance condition, baseline correction was performed in the power domain:
\[
V_{\mathrm{res}}=\sqrt{\max\!\left(V_{\mathrm{ON}}^{2}-V_{\mathrm{OFF}}^{2},\,0\right)},
\]
where \(V_{\mathrm{ON}}\) and \(V_{\mathrm{OFF}}\) are the 50~Hz RMS amplitudes measured in the generator-ON and generator-OFF recordings, respectively. The reported dB values were expressed as effective common-mode attenuation, computed as
\[
A_{\mathrm{CM}}=20\log_{10}\!\left(\frac{V_{\mathrm{inj}}}{V_{\mathrm{res}}}\right),
\]
where \(V_{\mathrm{inj}}=V_{\mathrm{pp}}/(2\sqrt{2})\) is the RMS amplitude corresponding to the nominal generator output of 1~V\(_{\mathrm{pp}}\). With this definition, larger positive values indicate stronger rejection. The resulting values were summarized on a per-channel basis for both balanced and mismatch conditions.

\subsection{Closed-Loop SSVEP Validation Protocol}
\label{subsec:closed_loop_ssvep_validation_protocol}

Closed-loop SSVEP validation was conducted to assess end-to-end operation of the embedded EEG platform under practical real-time use. Ten healthy participants aged 28--35 years with normal or corrected-to-normal vision took part in the online experiments. All recordings were conducted at the Computational Intelligence and Brain-Computer Interfaces Lab, University of Technology Sydney. Participants had no known history of neurological or psychiatric conditions and provided informed consent prior to participation. The study was conducted under ethics approval from the University of Technology Sydney Human Research Ethics Committee (UTS HREC REF No.~ETH20-5371).

\begin{figure}[ht]
\centering
\begin{subfigure}[t]{\linewidth}
    \centering
    \includegraphics[width=\linewidth]{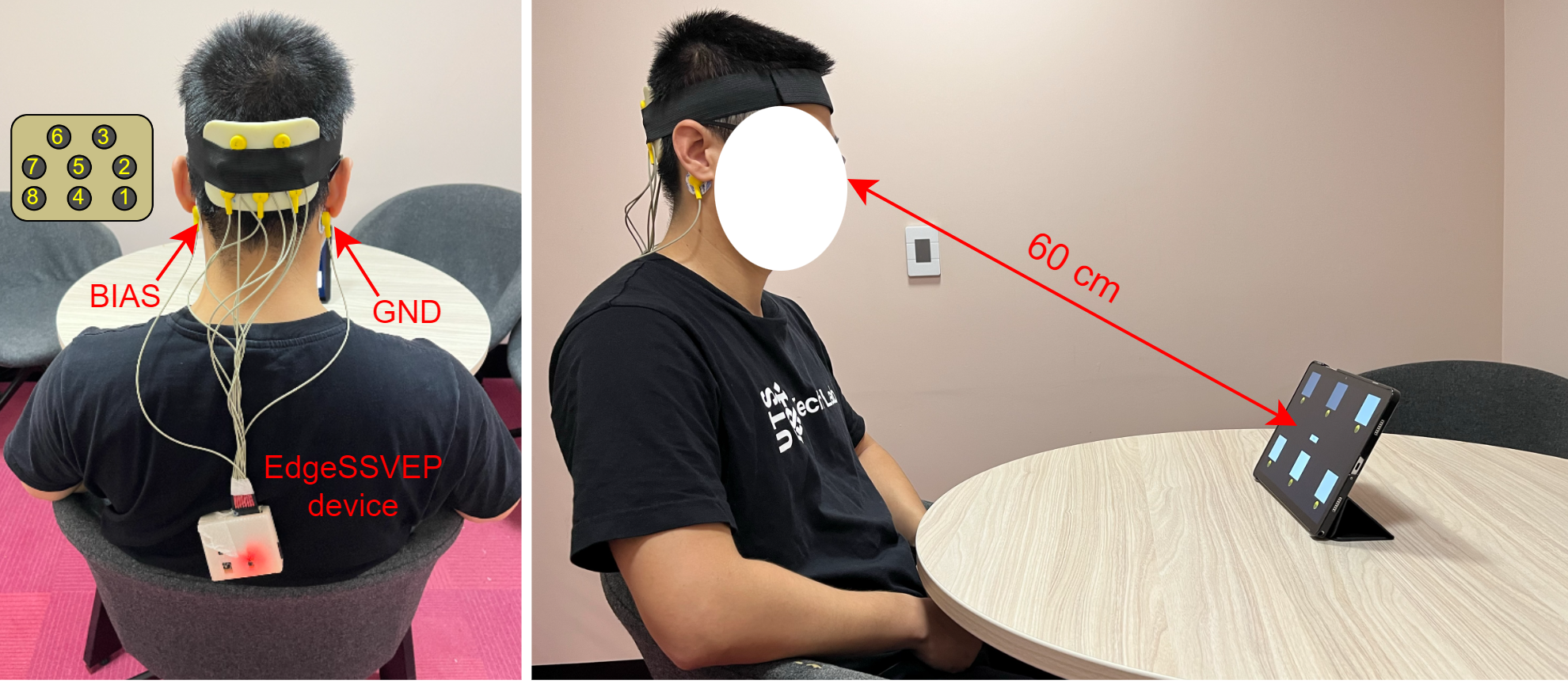}
    \caption{Experimental arrangement, in which the participant wore the occipital EEG sensor pad and viewed the six-target flicker interface on a tablet.}
    \label{fig:ssvep_experimental}
\end{subfigure}

\vspace{1mm}

\begin{subfigure}[t]{\linewidth}
    \centering
    \includegraphics[width=\linewidth]{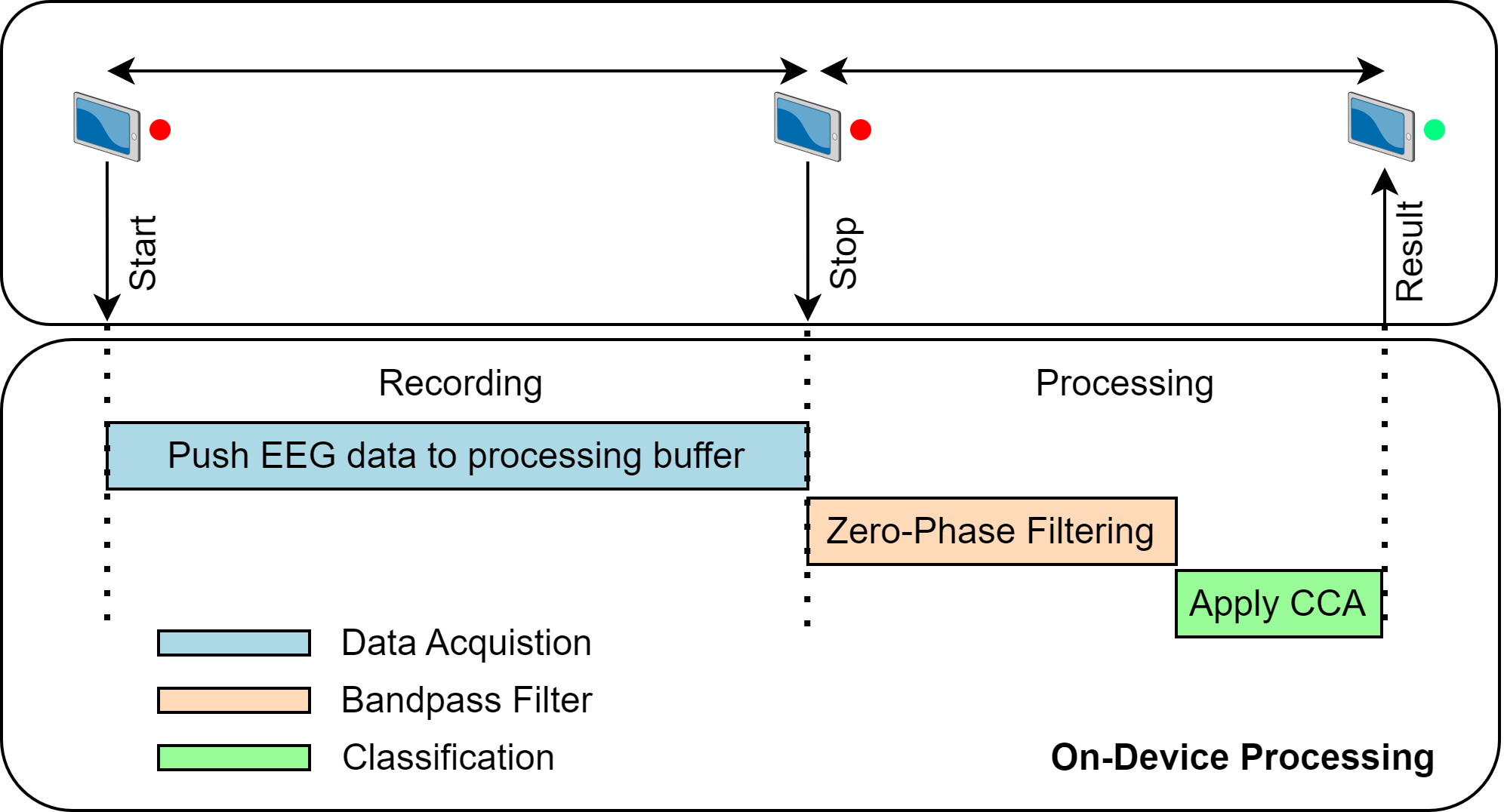}
    \caption{Trial-level closed-loop operation. The tablet issues a start event, EEG is acquired for 5~seconds, the final 4-second segment is processed on-device by the zero-phase bandpass filter and CCA classifier, and the detected target is returned to the tablet for real-time visual feedback.}
    \label{fig:ssvep_on_device_processing}
\end{subfigure}

\caption{Closed-loop SSVEP validation setup and operation flow.}
\label{fig:closed_loop_setup}
\end{figure}

Visual stimulation was presented on a Samsung Galaxy Tab A9+ with an 11-inch, 90~Hz display using a custom mobile application, as illustrated in Figure~\ref{fig:ssvep_experimental}. The interface presented six square flickering stimuli arranged in a \(2 \times 3\) grid, with target frequencies of 7, 8, 9, 11, 7.5, and 8.5~Hz. These stimulation frequencies were selected to remain outside the dominant spontaneous alpha band, which typically peaks around 10~Hz in occipital regions \cite{KLIMESCH1999169}. Frequencies close to the alpha range can be more difficult to separate from endogenous activity, whereas frequencies away from this region provide clearer spectral distinction for CCA-based recognition while remaining compatible with the 90~Hz refresh rate of the tablet display. During the experiment, the participant was seated approximately 60~cm from the display while wearing the occipital EEG sensor pad. Each stimulus had physical dimensions of \(3.21 \times 3.21\)~cm, corresponding to a visual angle of approximately \(3^\circ\) at this viewing distance. This arrangement kept the stimuli within the central visual field and promoted reliable SSVEP elicitation through foveal engagement \cite{yoshida-2020}.

The trial-level closed-loop operation is illustrated in Figure~\ref{fig:ssvep_on_device_processing}. At the beginning of each trial, the tablet transmitted a start event to the embedded device to initiate EEG acquisition, and a stop event was issued at the end of the acquisition period. Each participant completed 24 online trials. During each trial, visual flickers were presented for 5~seconds. The embedded system was configured to use the final 4-second segment of each 5-second trial as the analysis epoch for zero-phase bandpass filtering and CCA-based classification. This design choice was intended to reduce the influence of trial-onset transients and allow visual fixation to stabilize before decoding. Supporting evidence for this temporal selection is reported later in Table~\ref{tab:window_comparison}.

\section{Characterization and Validation Results}
\label{sec:characterization_validation_results}

This section presents the quantitative characterization and validation results of the embedded EEG platform. The reported results cover noise floor repeatability, sampling timing integrity, numerical fidelity relative to an offline double-precision reference, processing latency distribution, effective common-mode attenuation, power consumption, implementation footprint, and closed-loop SSVEP validation. These results assess the platform as an embedded measurement and processing system under both controlled characterization and practical end-to-end operating conditions.

\subsection{Noise Floor Repeatability}
\label{subsec:noise_floor_repeatability}

The shorted-input noise results are summarized in Table~\ref{tab:noise_repeatability}. Across six runs per operating mode, both the band-limited and wideband noise metrics remained highly stable under baseline and streaming conditions. In the 2--45~Hz band, which matches the embedded signal-processing pipeline, the mean RMS noise was \(0.0803 \pm 0.0007~\mu\mathrm{V}_{\mathrm{RMS}}\) in the OFF mode and \(0.0801 \pm 0.0004~\mu\mathrm{V}_{\mathrm{RMS}}\) in the ON mode. The corresponding wideband RMS values after DC removal were \(0.1534 \pm 0.0017~\mu\mathrm{V}_{\mathrm{RMS}}\) and \(0.1530 \pm 0.0006~\mu\mathrm{V}_{\mathrm{RMS}}\), respectively.

In both metrics, the ON-mode mean was marginally lower than the OFF-mode mean; however, the absolute differences were very small and remained within the observed across-run variation. This indicates that sustained TCP-based streaming did not produce any clear change in the system-level front-end noise floor within the resolution of the present experiment. The small across-run standard deviations further indicate good repeatability under both operating conditions.

\begin{table}[!t]
\caption{Noise floor repeatability using shorted inputs at 500~SPS (\(N=6\) runs per mode). Values are reported as mean \(\pm\) standard deviation across runs.}
\label{tab:noise_repeatability}
\centering
\setlength{\tabcolsep}{4pt}
\renewcommand{\arraystretch}{1.15}
\begin{tabular}{@{}lcc@{}}
\toprule
\textbf{Metric (\(\mu\)V\(_{\mathrm{RMS}}\))} & \textbf{OFF} & \textbf{ON} \\
\midrule
EEG-band (2--45~Hz) & \(0.0803 \pm 0.0007\) & \(0.0801 \pm 0.0004\) \\
Wideband (DC-removed) & \(0.1534 \pm 0.0017\) & \(0.1530 \pm 0.0006\) \\
\bottomrule
\end{tabular}
\end{table}

\subsection{Sampling Jitter and Drift}
\label{subsec:sampling_jitter_drift}

\begin{figure}[!t]
\centering
\includegraphics[width=\linewidth]{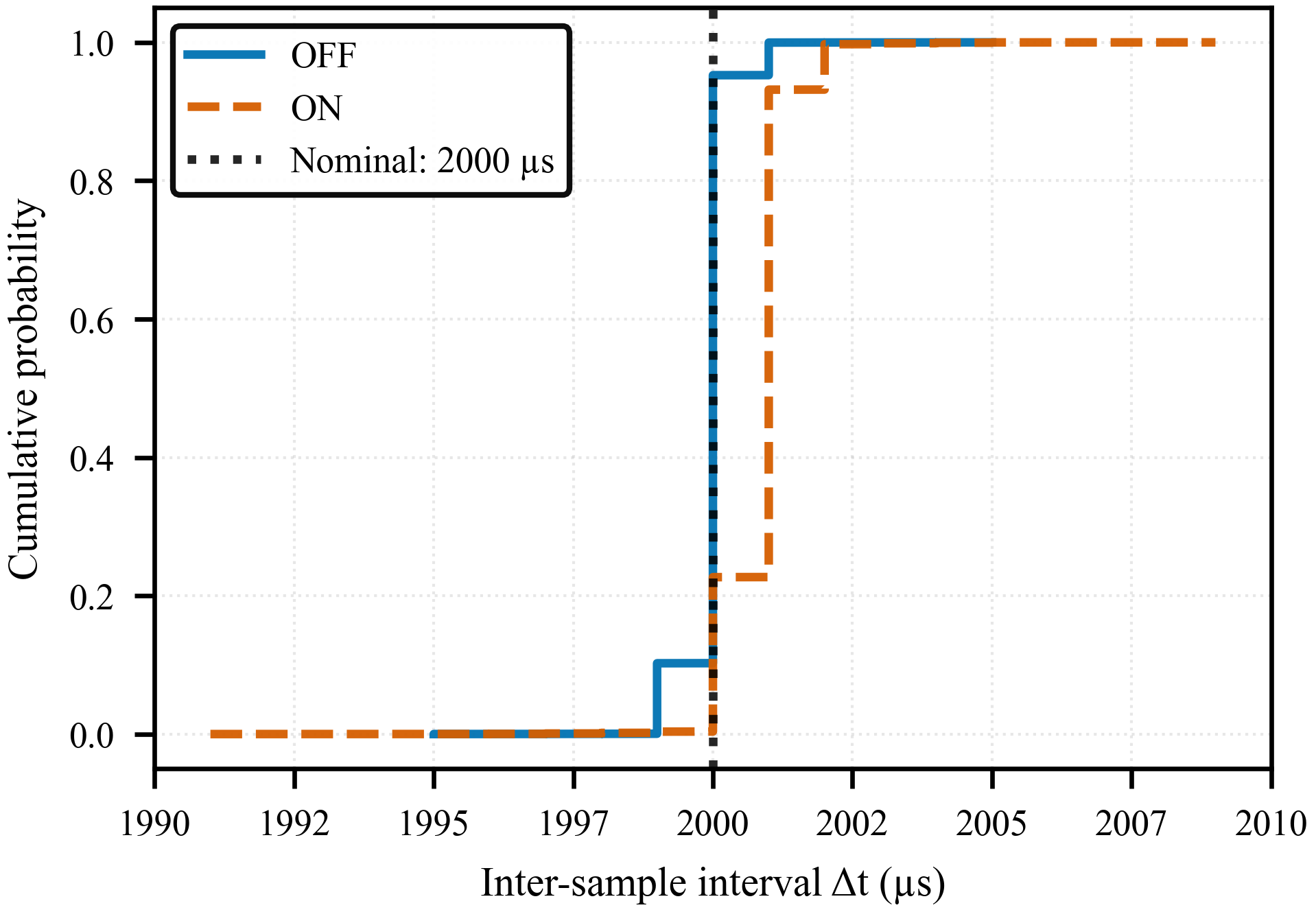}
\caption{Cumulative distribution of inter-sample intervals measured from ADS1299 DRDY timestamps under OFF and ON operating modes. The dotted vertical line indicates the nominal 2000~\(\mu\)s sampling period at 500~SPS.}
\label{fig:jitter_cdf_combined}
\end{figure}

The sampling timing results are summarized in Table~\ref{tab:sampling_jitter_drift} and Figure~\ref{fig:jitter_cdf_combined}. Across 300{,}000 recorded DRDY events per operating mode, the measured inter-sample interval remained tightly clustered around the nominal 2000~\(\mu\)s period in both OFF and ON conditions. In the OFF mode, the mean interval was 1999.94~\(\mu\)s with a standard deviation of 0.39~\(\mu\)s, whereas in the ON mode the mean interval was 2000.84~\(\mu\)s with a standard deviation of 0.56~\(\mu\)s.

The ON mode showed a small increase in both the mean interval and its spread; however, the absolute change remained very small relative to the nominal 2000~\(\mu\)s sampling period. Upper-tail statistics were similarly well controlled, with the 99th percentile remaining at 2001~\(\mu\)s in the OFF mode and 2002~\(\mu\)s in the ON mode. The minimum and maximum observed intervals were 1995 and 2005~\(\mu\)s in the OFF mode, and 1991 and 2009~\(\mu\)s in the ON mode.

Long-term drift remained negligible in both cases, at 0.42~ppm in the OFF mode and 0.89~ppm in the ON mode. These results show that DRDY-driven sampling remained highly deterministic and was not materially affected by sustained streaming within the resolution of the present experiment.

\begin{table}[!t]
\caption{Sampling jitter and drift at 500~SPS using 300{,}000 DRDY events per mode.}
\label{tab:sampling_jitter_drift}
\centering
\setlength{\tabcolsep}{4pt}
\renewcommand{\arraystretch}{1.15}
\begin{tabular}{@{}lcccccccc@{}}
\toprule
\textbf{Mode} & \textbf{Mean} & \textbf{Std} & \textbf{P50} & \textbf{P95} & \textbf{P99} & \textbf{Min} & \textbf{Max} & \textbf{Drift} \\
 & (\(\mu\)s) & (\(\mu\)s) & (\(\mu\)s) & (\(\mu\)s) & (\(\mu\)s) & (\(\mu\)s) & (\(\mu\)s) & (ppm) \\
\midrule
OFF & 1999.94 & 0.39 & 2000 & 2000 & 2001 & 1995 & 2005 & 0.42 \\
ON  & 2000.84 & 0.56 & 2001 & 2002 & 2002 & 1991 & 2009 & 0.89 \\
\bottomrule
\end{tabular}
\end{table}

\subsection{Numerical Fidelity}
\label{subsec:numerical_fidelity}

The numerical fidelity results are summarized in Table~\ref{tab:numerical_fidelity}. The embedded-equivalent configuration DF, which uses double-precision filtering and single-precision CCA, matched the full double-precision reference DD on all 240 trials, yielding 100\% decision agreement. In addition, the maximum absolute deviation in decision margin relative to DD was only \(1.02\times10^{-5}\), indicating that the use of single-precision arithmetic in the CCA stage introduced negligible numerical error in the present implementation.

In contrast, the configurations involving single-precision filtering showed lower fidelity relative to DD. Both FD and FF achieved 97.5\% agreement with DD, corresponding to 6 disagreements out of 240 trials, and both exhibited substantially larger decision-margin deviations than DF. The identical behavior of FD and FF indicates that the dominant source of numerical deviation arose from the reduced precision of the filtering stage rather than the CCA stage.

These results show that the deployed mixed-precision embedded pipeline preserved the decision behavior of the double-precision reference, and that single-precision CCA was not the limiting factor in numerical fidelity for the present implementation. This mixed-precision configuration, consisting of double-precision zero-phase filtering and single-precision CCA, was also the arithmetic used in the embedded platform for the latency measurements reported next.

\begin{table}[!t]
\centering
\caption{Numerical fidelity relative to the double-precision reference (DD) over 240 trials.}
\label{tab:numerical_fidelity}
\begin{tabular}{@{}lccc@{}}
\toprule
\textbf{Configuration vs DD} & \textbf{Agreement (\%)} & \textbf{Max \(|\Delta m|\)} & \textbf{Diff.} \\
\midrule
DF (double/float) & 100.0 & \(1.02\times10^{-5}\) & 0/240 \\
FD (float/double) & 97.5  & 0.182                 & 6/240 \\
FF (float/float)  & 97.5  & 0.182                 & 6/240 \\
\bottomrule
\end{tabular}

\vspace{1mm}
\begin{tablenotes}
\footnotesize
\item \(m\) denotes the decision margin, defined as \(m=\rho_{\mathrm{peak}}-\rho_{\mathrm{2nd}}\). \\The reported deviation is \(\Delta m=m_{\mathrm{cfg}}-m_{\mathrm{DD}}\).
\end{tablenotes}
\end{table}

\subsection{Processing Latency Distribution}
\label{subsec:processing_latency_distribution}

\begin{figure}[!t]
\centering
\includegraphics[width=\linewidth]{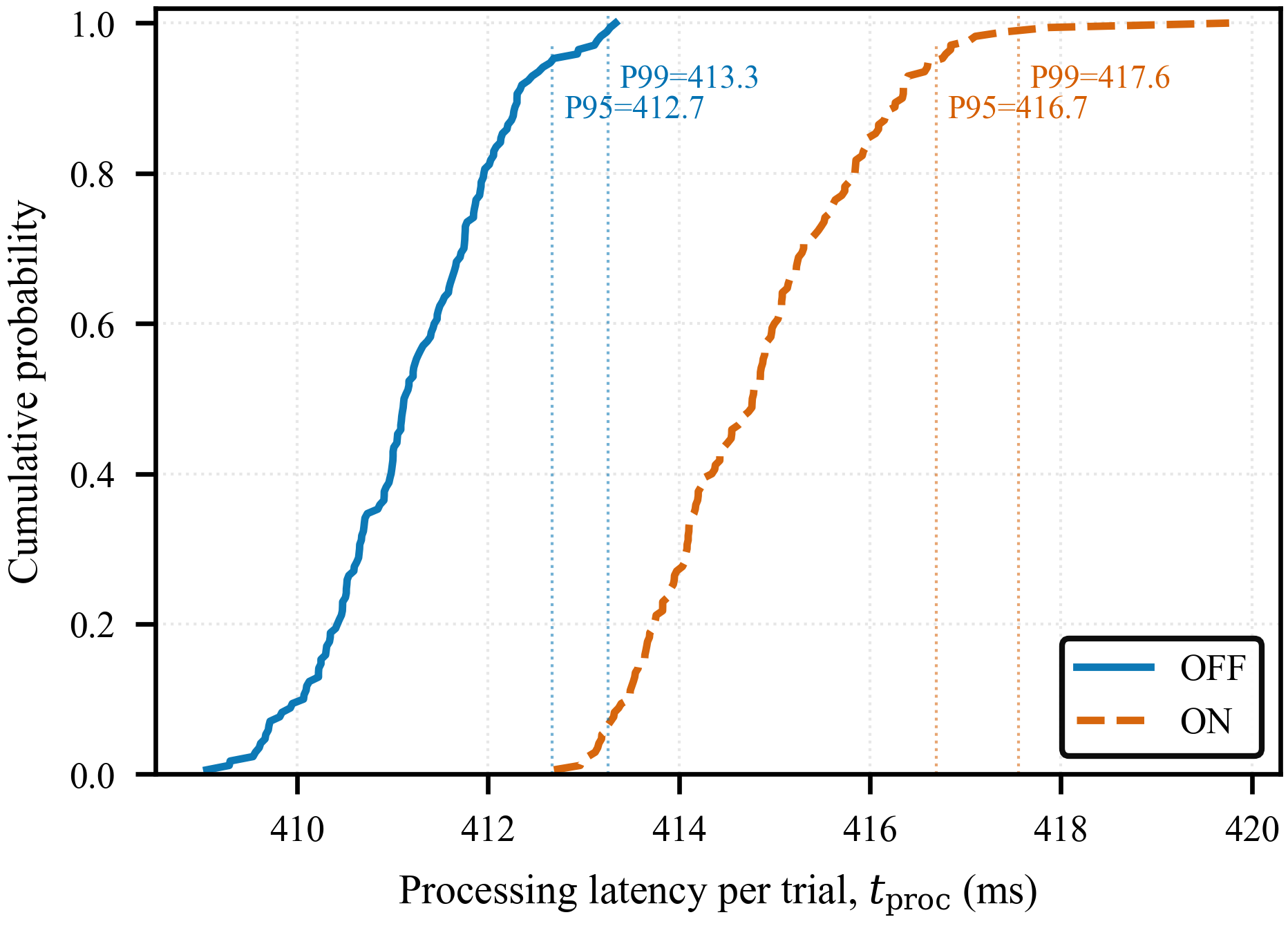}
\caption{Cumulative distribution of processing latency per trial under OFF and ON operating modes. Vertical dotted lines indicate the 95th and 99th percentile values for each mode.}
\label{fig:processing_latency_cdf}
\end{figure}

The processing latency results correspond to the deployed mixed-precision embedded pipeline, which uses double-precision zero-phase filtering and single-precision CCA. Table~\ref{tab:processing_latency} and Figure~\ref{fig:processing_latency_cdf} summarize the resulting latency statistics. Across 170 processing cycles per operating mode, the end-to-end latency from the start of zero-phase filtering to the completion of the CCA decision remained tightly distributed in both OFF and ON conditions. The mean latency was 411.20~ms in the OFF mode and 414.81~ms in the ON mode, corresponding to an increase of 3.61~ms under streaming. Despite this increase, the total latency remained well below the 5-second trial duration, leaving substantial timing margin for real-time operation.

The ON mode showed slightly higher latency and variability than the OFF mode, but the overall spread remained small. The standard deviation was 0.91~ms in the OFF mode and 1.14~ms in the ON mode, while the 99th percentile remained below 418~ms in both cases. The gap between the mean and the 99th percentile was 2.05~ms in the OFF mode and 2.74~ms in the ON mode, indicating tightly bounded execution time without pronounced outliers.

Stage-level analysis showed that the zero-phase bandpass filter accounted for the larger share of the total processing time, with mean latencies of 273.12~ms in the OFF mode and 275.05~ms in the ON mode, whereas the CCA stage contributed 138.08~ms and 139.76~ms, respectively. Overall, the embedded pipeline operated deterministically, with concurrent streaming adding only a small timing overhead.

\begin{table}[!t]
\caption{Processing latency distribution for end-to-end on-device decoding, using 170 processing cycles per mode.}
\label{tab:processing_latency}
\centering
\setlength{\tabcolsep}{4pt}
\renewcommand{\arraystretch}{1.15}
\begin{tabular}{@{}lcccccc@{}}
\toprule
\textbf{Mode} & \textbf{Mean} & \textbf{Std} & \textbf{P50} & \textbf{P95} & \textbf{P99} & \textbf{Max} \\
 & (ms) & (ms) & (ms) & (ms) & (ms) & (ms) \\
\midrule
OFF & 411.20 & 0.91 & 411.13 & 412.67 & 413.25 & 413.33 \\
ON  & 414.81 & 1.14 & 414.77 & 416.69 & 417.55 & 419.76 \\
\bottomrule
\end{tabular}
\end{table}

\subsection{Effective Common-Mode Rejection}
\label{subsec:effective_common_mode_rejection}

\begin{figure}[!t]
\centering
\includegraphics[width=\linewidth]{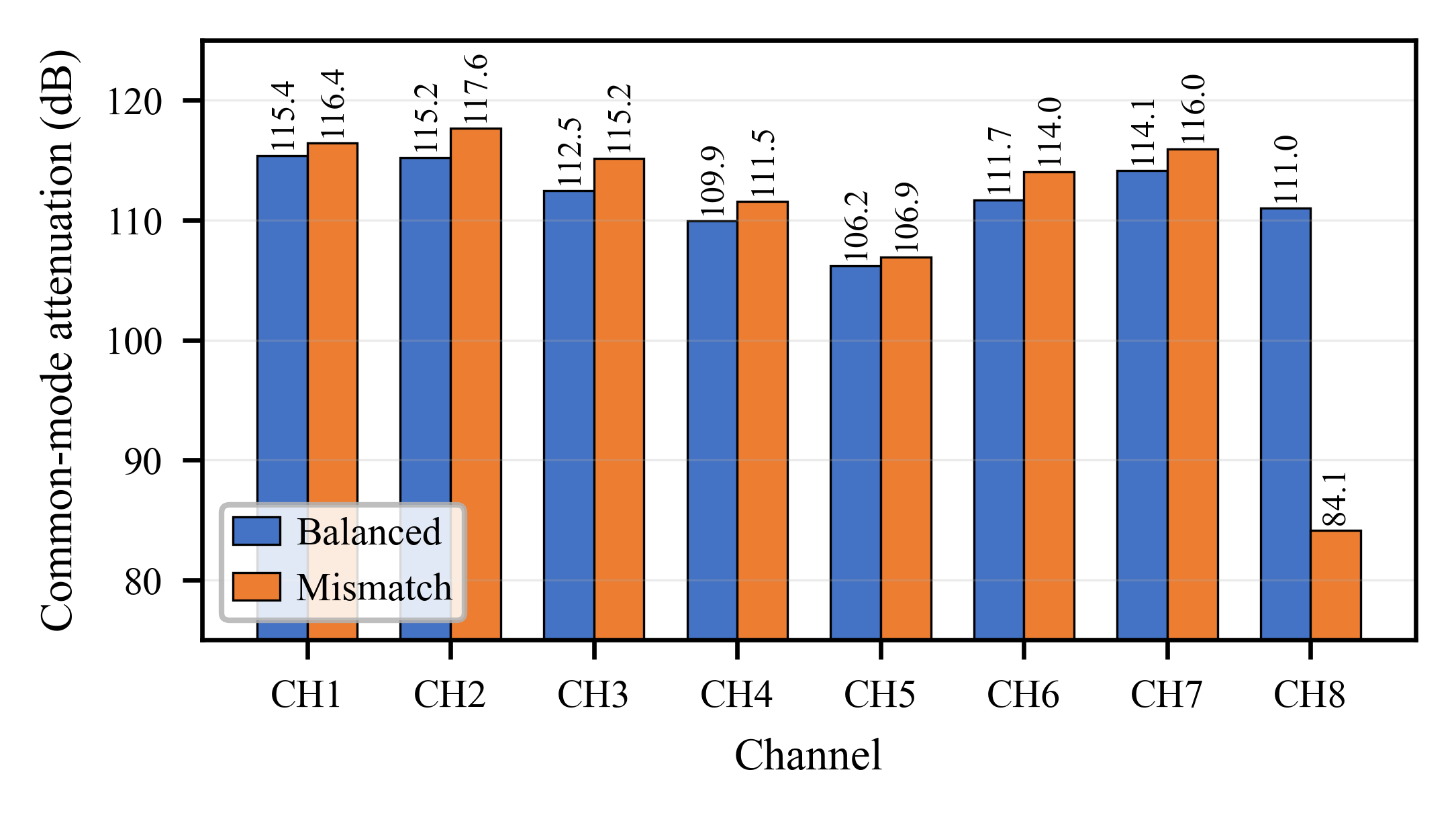}
\caption{Channel-wise effective common-mode attenuation under balanced and mismatch source-impedance conditions. The mismatch case introduces a 20~k\(\Omega\) source impedance on CH8, while the remaining channels remain at 10~k\(\Omega\).}
\label{fig:cmrr_bar}
\end{figure}

The effective common-mode attenuation results are summarized in Table~\ref{tab:cmrr_results} and Figure~\ref{fig:cmrr_bar}. Under the balanced condition, the measured attenuation ranged from 106.2 to 115.4~dB across the eight channels, with a median of 112.1~dB. These values indicate strong suppression of the injected 50~Hz common-mode component under matched source-impedance conditions.

Under the mismatch condition, Channels~1--7 showed only small changes relative to the balanced case, with attenuation shifts ranging from +0.7 to +2.7~dB. In contrast, the intentionally perturbed channel, CH8, exhibited a pronounced reduction from 111.0 to 84.1~dB, corresponding to a degradation of 26.9~dB. Thus, the mismatch condition did not produce a uniform reduction across all channels; instead, the dominant effect was a localized loss of common-mode attenuation at the intentionally unbalanced channel.

Small changes were also observed on Channels~1--7 despite their unchanged source resistances. Because the channels share the same common injection and reference network, the imposed asymmetry at CH8 can slightly redistribute the residual common-mode coupling seen by the remaining channels. These shifts were minor relative to the localized degradation at CH8 and should not be interpreted as a global improvement in rejection performance.

Although the median attenuation increased slightly from 112.1 to 114.6~dB, this shift was driven by the small upward changes observed in the non-perturbed channels rather than by improved rejection at the mismatched input. Overall, the channel-wise results show that the imposed source-impedance asymmetry primarily affected the mismatched path, while the remaining channels stayed close to their balanced-condition behavior. These results show that system-level common-mode attenuation remained robust under balanced operation, but was sensitive to localized source-impedance mismatch.

\begin{table}[!t]
\caption{Effective common-mode attenuation at 50~Hz under balanced and mismatch source-impedance conditions.}
\label{tab:cmrr_results}
\centering
\setlength{\tabcolsep}{4pt}
\renewcommand{\arraystretch}{1.15}
\begin{tabular}{@{}lccc@{}}
\toprule
\textbf{Channel} & \textbf{Balanced (dB)} & \textbf{Mismatch (dB)} & \textbf{\(\Delta\) (dB)} \\
\midrule
CH1 & 115.4 & 116.4 & +1.0 \\
CH2 & 115.2 & 117.6 & +2.4 \\
CH3 & 112.5 & 115.2 & +2.7 \\
CH4 & 109.9 & 111.5 & +1.6 \\
CH5 & 106.2 & 106.9 & +0.7 \\
CH6 & 111.7 & 114.0 & +2.3 \\
CH7 & 114.1 & 116.0 & +1.9 \\
CH8 & 111.0 & 84.1  & -26.9 \\
\midrule
Median & 112.1 & 114.6 & +2.5 \\
\bottomrule
\end{tabular}
\end{table}

\subsection{Power Consumption and Implementation Footprint}
\label{subsec:power_consumption_implementation_footprint}

The power and memory-footprint results are summarized in Tables~\ref{tab:power_freq} and \ref{tab:implementation_footprint}. These resource measurements are particularly relevant because they correspond to the actual deployed embedded pipeline used in the characterization and closed-loop validation experiments, rather than to a simplified prototype or a partially offloaded implementation. Power consumption was evaluated at CPU frequencies of 80, 160, and 240~MHz under two representative operating modes. The OFF mode corresponds to the baseline condition without external EEG streaming, whereas the ON mode corresponds to the streaming-enabled condition in which raw EEG data are transmitted over Wi-Fi to an external client.

In the OFF mode, the embedded system operated autonomously, performing real-time EEG acquisition, zero-phase bandpass filtering, and CCA-based classification entirely on-device, while transmitting only the classification result to the connected mobile application over a TCP socket. In the ON mode, the same processing pipeline remained active, but raw EEG data were additionally streamed over Wi-Fi to the connected client. This dual operation increased power consumption primarily because of the continuous wireless transmission.

\begin{table}[!t]
\centering
\begin{threeparttable}
\caption{Power profile versus CPU frequency in OFF and ON Wi-Fi modes.}
\label{tab:power_freq}
\scriptsize
\begin{tabular}{@{}lcccccc@{}}
\toprule
\multirow{2}{*}{\textbf{Component}} &
\multicolumn{2}{c}{\textbf{240\,MHz}} &
\multicolumn{2}{c}{\textbf{160\,MHz}} &
\multicolumn{2}{c}{\textbf{80\,MHz}} \\
\cmidrule(lr){2-3}\cmidrule(lr){4-5}\cmidrule(lr){6-7}
& OFF & ON & OFF & ON & OFF & ON \\
\midrule
ESP32-S3 (core)\textsuperscript{\textdagger} & 132.0 & 132.0 & 112.2 & 112.2 & 95.7 & 95.7 \\
Wi-Fi (AP)\textsuperscript{\textdaggerdbl}   & 26.4  & 138.6 & 26.4  & 132.0 & 26.4 & 82.5 \\
ADS1299\textsuperscript{\textsection}        & 42.0  & 42.0  & 42.0  & 42.0  & 42.0 & 42.0 \\
Others (LDO, LED)                            & 22.0  & 22.0  & 22.0  & 22.0  & 22.0 & 22.0 \\
\midrule
\textbf{Total (mW)} & \textbf{222.4} & \textbf{334.6} &
\textbf{202.6} & \textbf{308.2} &
\textbf{186.1} & \textbf{242.2} \\
\bottomrule
\end{tabular}

\vspace{1mm}
\begin{tablenotes}
\small
\item \textsuperscript{\textdagger} Baseline ESP32-S3 power running the EEG pipeline (SPI, filtering, CCA) with Wi-Fi/BLE disabled.
\item \textsuperscript{\textdaggerdbl} Additional power when the Wi-Fi subsystem is enabled.
\item \textsuperscript{\textsection} ADS1299 at 8 channels, 500\,SPS, PGA=12, internal clock. Estimated 42\,mW at AVDD-AVSS=5\,V and DVDD=3.3\,V.
\end{tablenotes}
\end{threeparttable}
\end{table}

Table~\ref{tab:power_freq} shows that total power increased with CPU frequency in both modes and was consistently higher in the ON mode because of the additional wireless transmission overhead. At 240~MHz, the total power increased from 222.4~mW in the OFF mode to 334.6~mW in the ON mode. At 160~MHz, the corresponding values were 202.6 and 308.2~mW, whereas at 80~MHz they were 186.1 and 242.2~mW. These results indicate that Wi-Fi activity dominates the incremental power cost, while the embedded signal-processing pipeline itself remains compatible with battery-powered operation.

The memory footprint of the deployed signal-processing pipeline is summarized in Table~\ref{tab:implementation_footprint}. The reported memory usage corresponds to the deployed operating configuration, in which the final 4-second EEG segment of each 5-second trial is processed at 500~SPS, yielding 2000 samples per channel for the zero-phase bandpass filter and CCA classifier. Under this configuration, the zero-phase bandpass filter uses three PSRAM work buffers totaling approximately 96~kB. The CCA stage allocates its large sample and transpose buffers in PSRAM with an additional footprint of approximately 192~kB, while smaller CCA matrices occupy about 1.2~kB in internal SRAM. In addition, the precomputed sinusoidal reference bank requires approximately 192~kB of read-only storage. This allocation strategy places the largest arrays in external memory while reserving internal SRAM for latency-sensitive matrix operations, confirming that the full mixed-precision pipeline fits within the ESP32-S3 memory hierarchy without external compute support.

\begin{table}[!t]
\caption{Memory footprint of the deployed signal-processing pipeline for the 4-second, 500~SPS analysis window.}
\label{tab:implementation_footprint}
\centering
\setlength{\tabcolsep}{4pt}
\renewcommand{\arraystretch}{1.15}
\begin{tabular}{@{}lcc@{}}
\toprule
\textbf{Component} & \textbf{Memory type} & \textbf{Footprint} \\
\midrule
BPF working buffers & PSRAM & 96~kB \\
CCA sample / transpose buffers & PSRAM & 192~kB \\
CCA small matrices & Internal SRAM & 1.2~kB \\
Sinusoidal reference bank & Read-only memory & 192~kB \\
\bottomrule
\end{tabular}
\end{table}

\subsection{Closed-Loop SSVEP Validation}
\label{subsec:closed_loop_ssvep_validation}

The closed-loop validation results are summarized in Tables~\ref{tab:closed_loop_results} and \ref{tab:window_comparison}. Across 10 participants and 24 online trials per participant, the embedded platform achieved an overall classification accuracy of 99.17\%, corresponding to 238 correct detections out of 240 trials. These results were obtained in the ON mode using the deployed embedded pipeline. Subject-wise performance remained consistently high, with eight participants achieving 100\% accuracy and the remaining two participants achieving 95.83\%. This indicates reliable end-to-end operation under practical closed-loop conditions, including EEG acquisition, embedded decoding, Wi-Fi communication, and real-time visual feedback on the tablet.

Table~\ref{tab:window_comparison} provides supporting analysis for the deployed temporal-window selection by comparing alternative analysis windows using the recorded online trials. This comparison was performed offline using the same CCA-based evaluation pipeline corresponding to the deployed embedded system. Among the tested options, the final 4-second interval achieved the highest accuracy at 99.17\%, compared with 94.17\% for the first 4-second interval and 97.92\% for the full 5-second interval. These results support the use of the final 4-second segment in the embedded pipeline.

Because the online validation used the same deployed pipeline characterized in Section~\ref{subsec:processing_latency_distribution}, the relevant decision-time behavior is represented by the ON-mode latency results reported there rather than by the earlier 5-second offline processing estimate. The information transfer rate (ITR) was computed using the standard formulation for multi-class BCI communication rate~\cite{wolpaw-2002}:
\[
\mathrm{ITR} = \left[\log_2 M + P\log_2 P + (1-P)\log_2\!\left(\frac{1-P}{M-1}\right)\right]\frac{60}{T},
\]
where \(M=6\) is the number of targets, \(P\) is the online classification accuracy, and \(T\) is the average time required for one selection. Using the measured online accuracy of 99.17\% and the ON-mode end-to-end selection time, the resulting ITR was 27.66~bits/min. These results confirm that the platform supports both quantitative system characterization and practical closed-loop SSVEP operation.

\begin{table}[!t]
\caption{Closed-loop online classification results for 10 subjects using the embedded SSVEP BCI system in the ON mode.}
\label{tab:closed_loop_results}
\centering
\setlength{\tabcolsep}{4pt}
\renewcommand{\arraystretch}{1.15}
\begin{tabular}{@{}ccc@{}}
\toprule
\textbf{Subj.} & \textbf{Correct / 24} & \textbf{Acc. (\%)} \\
\midrule
S01 & 24/24 & 100.00 \\
S02 & 24/24 & 100.00 \\
S03 & 24/24 & 100.00 \\
S04 & 24/24 & 100.00 \\
S05 & 23/24 & 95.83 \\
S06 & 24/24 & 100.00 \\
S07 & 24/24 & 100.00 \\
S08 & 24/24 & 100.00 \\
S09 & 24/24 & 100.00 \\
S10 & 23/24 & 95.83 \\
\midrule
\textbf{Avg.} & \textbf{238/240} & \textbf{99.17 \(\pm\) 1.76} \\
\bottomrule
\end{tabular}
\end{table}

\begin{table}[!t]
\caption{Offline comparison of temporal analysis windows using the recorded online trials and the same CCA-based evaluation pipeline as the deployed system.}
\label{tab:window_comparison}
\centering
\setlength{\tabcolsep}{4pt}
\renewcommand{\arraystretch}{1.15}
\begin{tabular}{@{}lcc@{}}
\toprule
\textbf{Window} & \textbf{Correct / Total} & \textbf{Accuracy (\%)} \\
\midrule
First 4~s & 226 / 240 & 94.17 \\
Final 4~s & 238 / 240 & 99.17 \\
Full 5~s  & 235 / 240 & 97.92 \\
\bottomrule
\end{tabular}
\end{table}

\section{Discussion}
\label{sec:discussion}

\subsection{Measurement-Oriented Interpretation of System Performance}
\label{subsec:measurement_oriented_interpretation}

The characterization results support interpreting the proposed platform not only as an embedded SSVEP decoder, but also as a quantitatively evaluated EEG measurement and processing instrument. The reported metrics cover complementary aspects of system trustworthiness: front-end stability through shorted-input noise, temporal integrity through sampling jitter and drift, computational faithfulness through numerical-fidelity analysis, deterministic execution through latency distribution, and interference robustness through effective common-mode attenuation.

The OFF/ON comparison is particularly important because it shows that enabling continuous telemetry did not materially disturb the core measurement path. Front-end noise remained stable, inter-sample timing stayed tightly bounded with sub-microsecond jitter and negligible drift, and processing latency increased only slightly under streaming. The numerical-fidelity results further show that the deployed mixed-precision pipeline preserved the decision behavior of the double-precision reference, while the mismatch experiment provided a physically interpretable view of sensitivity to localized source-impedance imbalance.

Overall, the results show that the platform combines embedded real-time operation with quantitative evidence of signal-quality stability, timing integrity, computational fidelity, and interference robustness under realistic operating conditions.

\subsection{Design Trade-offs in Embedded Signal Processing}
\label{subsec:design_tradeoffs_embedded_filtering_cca}

The signal chain reflects a deliberate trade-off between signal integrity and computational cost. Zero-phase filtering was prioritized to prevent phase distortion, despite accounting for 66\% of total processing latency. This latency ($\approx 415$~ms) remains well within the real-time requirements for a 5-second trial structure.

The memory strategy optimizes the ESP32-S3 hierarchy by placing high-volume buffers (288~kB) in PSRAM while reserving internal SRAM for latency-sensitive matrix inversions. Power measurements indicate that the 222~mW baseline is dominated by the AFE and CPU, with Wi-Fi streaming adding a $\approx 112$~mW overhead. While not optimized for ultra-low-power like specialized SoCs, the system achieves a balanced operating point for autonomous, high-fidelity decoding.

\subsection{Comparison With Prior Embedded EEG and SSVEP Systems}
\label{subsec:comparison_prior_embedded_systems}

Table~\ref{tab:comparison_embedded_systems} places the proposed platform in the context of representative embedded and portable EEG/SSVEP systems. A direct metric-by-metric comparison is inherently difficult because many prior studies emphasize low power, portability, or classification performance, whereas comparatively fewer report instrumentation-oriented characterization such as shorted-input noise repeatability, sampling jitter and drift, numerical fidelity, processing-latency distribution, or system-level common-mode attenuation. This difference in reporting scope is itself one of the key observations arising from the comparison.

Among the most relevant embedded comparators, BioWolf~\cite{kartsch-2019} demonstrates an impressive ultra-low-power wearable architecture with 8 channels, embedded processing, online validation, and some electrical characterization. In that respect, it is a strong example of a highly optimized wearable BCI platform. Teversham \textit{et al.}~\cite{teversham-2022} provide a particularly close architectural comparison because they also demonstrate real-time on-device SSVEP decoding on an ESP32-class microcontroller. However, those works primarily emphasize low-power wearable optimization or low-cost embedded decoding, respectively, whereas the present paper emphasizes broader quantitative characterization of the embedded platform as a measurement and processing system.

EEG-Linx~\cite{ding-2025} and the multi-modal instrument of Barras \textit{et al.}~\cite{barras-2026} are informative in a different way. These systems illustrate that portable and wearable neurotechnology platforms can be designed with strong attention to synchronization, modularity, and real-time acquisition integrity. However, they are not presented as autonomous embedded SSVEP decoding systems with closed-loop validation. In contrast, the proposed platform combines multichannel embedded SSVEP decoding, closed-loop operation, timing characterization, numerical fidelity analysis, and effective common-mode attenuation characterization within a single microcontroller-class implementation.

The main distinction of this work is not simply its online accuracy, but the combination of embedded real-time operation with a broader measurement-oriented evaluation framework. In this sense, the paper contributes not only a device, but also a reference style of characterization for future embedded EEG/SSVEP systems, where timing integrity, arithmetic fidelity, and interference robustness are reported alongside task-level decoding performance.

\begin{table*}[ht]
\centering
\caption{Comparison with representative embedded and portable EEG/SSVEP systems. NR: not reported.}
\label{tab:comparison_embedded_systems}
\scriptsize
\setlength{\tabcolsep}{4pt}
\renewcommand{\arraystretch}{1.15}
\begin{tabularx}{\textwidth}{@{}lXc c c c c c c@{}}
\toprule
\textbf{System} & \textbf{Platform} & \textbf{Ch.} & \textbf{On-device decoding} & \textbf{Power (mW)} & \textbf{Timing / jitter} & \textbf{Numerical fidelity} & \textbf{CM attenuation} & \textbf{Closed-loop} \\
\midrule
BioWolf~\cite{kartsch-2019} &
Mr.~Wolf + Nordic BLE MCU + ADS1298 &
8 & Yes & 6.31 & No & No & Yes & Yes \\

Teversham et al.~\cite{teversham-2022} &
ESP32 (240~MHz) + custom single-channel EEG front end &
1 & Yes & NR & No & No & No & Yes \\

EEG-Linx~\cite{ding-2025} &
nRF52833 + ADS1299 wireless EEG sensor network &
4/node & No & NR & Yes & No & No & No \\

Barras et al.~\cite{barras-2026} &
STM32H7 + ESP32-S2 + ADS1299 multi-modal instrument &
32 & No & 280--840 & Yes & No & Yes & No \\

\midrule
\textbf{This work} &
ESP32-S3 (240~MHz) + ADS1299 &
8 & Yes & 222 & Yes & Yes & Yes & Yes \\
\bottomrule
\end{tabularx}
\end{table*}

\subsection{Limitations and Future Work}
\label{subsec:limitations_future_work}

The current implementation is restricted to CCA to ensure deterministic execution and training-free operation. While effective, future work could explore adaptive algorithms or deep learning classifiers as MCU-class AI accelerators evolve. Additionally, the common-mode experiment highlights a sensitivity to localized source-impedance mismatch ($26.9$~dB degradation at the perturbed channel), suggesting that future designs would benefit from active electrode shielding. Finally, extending this measurement framework to motion-intensive conditions and long-term stability testing remains a priority for wearable deployment.

\section{Conclusion}
\label{sec:conclusion}

This paper presented an embedded EEG instrumentation platform for real-time SSVEP decoding based on an ESP32-S3 microcontroller and an ADS1299 analog front end. The proposed system performs multichannel EEG acquisition, zero-phase bandpass filtering, and CCA-based classification entirely on-device, while supporting wireless communication and closed-loop operation.

A central contribution of this work is the quantitative characterization of the embedded platform from both instrumentation and implementation perspectives. The reported results showed stable shorted-input noise behavior under both baseline and streaming conditions, tightly bounded sampling jitter with negligible long-term drift, exact decision agreement between the deployed mixed-precision pipeline and the double-precision reference, deterministic processing latency well within practical real-time limits, and strong system-level effective common-mode attenuation under balanced input conditions. The source-impedance mismatch experiment further showed that the degradation in common-mode attenuation was localized to the intentionally perturbed channel, providing an interpretable system-level view of interface sensitivity. In addition, the power and memory-footprint measurements confirmed that the full deployed 4-second, 500~SPS analysis pipeline can be executed on a microcontroller-class device without external compute support.

The platform was further validated in a closed-loop six-target SSVEP experiment, where it achieved 99.17\% online accuracy and an information transfer rate of 27.66~bits/min in the ON mode. Together, these results show that the proposed system functions not only as an embedded SSVEP decoder, but also as a quantitatively characterized EEG measurement and processing platform for real-time operation.

Beyond the specific device presented here, the study also establishes a measurement-oriented evaluation framework for future embedded EEG/SSVEP systems, in which signal quality, timing integrity, computational fidelity, and interference robustness are assessed alongside task-level decoding performance.

\bibliographystyle{unsrt}
\balance
\bibliography{TIM-R0}

\end{document}